# Experimental Demonstration of 3D Reflected Beamforming at sub6GHz thanks to Varactor Based Reconfigurable Intelligent Surface

Philippe Ratajczak*, Eric Séguenot*, Dinh-Thuy Phan-Huy**
Orange Innovation/Networks
*Sophia-Antipolis, **Chatillon, France
*philippe.ratajczak@orange.com

*Abstract*—Reconfigurable intelligent surface (RIS) is a promising solution to boost coverage sustainably by reflecting waves from a transmitter to a receiver and acting as a low-power and passive relay. In this paper, for the first time, we demonstrate experimentally that a reconfigurable intelligent surface designed for sub6GHz, and using varactor technology, can perform three-dimensional reflective beamforming. This result is achieved with a RIS prototype of 984 unit-cells, thanks to a compact control circuit individually addressing and configuring the voltage of each unit-cell, with a distinct voltage. To our knowledge, this prototype configures 17 to 70 times more distinct voltages than in the state-of-the-art. The experimental results in an indoor environment show a 10 dB gain. They also show, for the first time, that producing such a new prototype is feasible with minimal energy footprint and environmental impact, thanks to refurbishing. Indeed, a reflectarray antenna originally designed for three-dimensional beamforming has been turned into a RIS.

## I. Introduction

Networks of the future 6th generation (6G) have the opportunity to improve the three societal values of trustworthiness, inclusion, and sustainability [1]. To increase the sustainability of future 6G networks, [2] has proposed to introduce a new type of node called reconfigurable intelligent surfaces (RIS) [2]. A reflective RIS [3] is a sort of intelligent mirror of waves that reflects impinging waves from a transmitter toward a target receiver. It is a kind of passive relay, with a potentially larger energy efficiency than an active relay [3]. RIS is therefore a promising sustainable (low-energy) technology to boost coverage, localization accuracy, secrecy, or Electromagnetic Field Exposure reduction [4].

Many different implementations of RIS exist. Recently, very few prototypes for sub6GHz bands based on varactors have been demonstrated successfully [5-9]. In these prototypes, a RIS is an array of unit-cell patch antennas. Each unit-cell reradiates the impinging wave with a distinct phase shift. Each phase-shift is controlled through varactors that translate a voltage into a phase shift. The advantage of such implementation is that a continuous phase-shift control is possible. However, to our knowledge, all experiments have been focused in demonstrating two-dimensional (2D) reflective beamforming (BF).

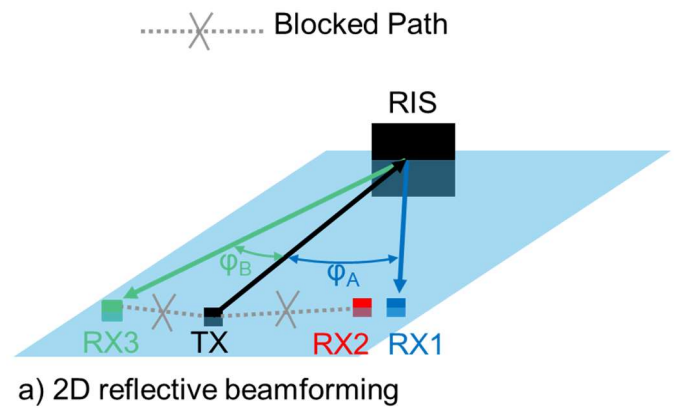

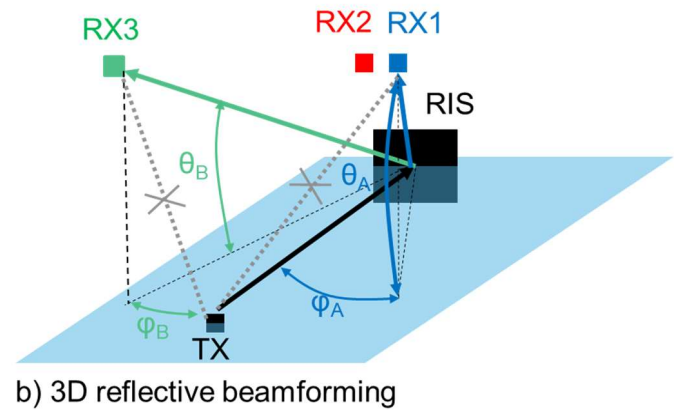

Fig. 1. 2D reflective beamforming versus 3D reflective beamforming

Fig. 1 illustrates the difference between 2D and three-dimensional (3D) reflected BF, in the case where a transmitter (TX) transmits towards a receiver 1 (RX1), or a receiver 2 (RX2) or a receiver 3 (RX3), RX1 and RX2 being close to each other.

A RIS is used to act as a passive relay between TX and RX1, RX2, or RX3. As illustrated in Fig. 1-a), in 2D reflected BF, all located in the same plan. In this case, to relay waves to RX1 (or RX2 close to RX1), the RIS tunes its unit-cells phase-shifts, according to the azimuth angle $\phi_A$ between the TX and RX1. Similarly, to relay waves to RX3, the RIS tunes its unit-cells phase-shifts, according to the angle $\phi_B$ between the TX and RX3. As illustrated by the example of Fig. 1-b), in 3D BF, the unit-cell has to tune the unit-cells phase-shifts according to additional angles, the elevation angles $\theta_A$ and $\theta_B$, to reach RX1 together with RX2, or RX3, respectively.

In previous RIS prototypes, 2D reflective BF implementation lowers the complexity of the RIS electronic control circuit. In [5,6], even though the RIS has an array of 196 unit-cells (14 columns and 14 lines), the same phase-shift is applied to the 14 unit-cells of an individual column; therefore, the RIS uses only 14 different phase-shifts. In [7], even though the RIS prototype has a square array of 2430 unit-cells, again, the control is per column, leading to around 50 phase-shifts. In [8], a meta-surface with a fixed configuration. In [9], though the array has 20 rows and 55 columns (i.e., 1,100 unit cells), the reflected BF is experimentally demonstrated in the azimuth domain only, i.e., column-wise.

In this paper, for the first time, we present experimental results showing the ability of a varactor-based RIS prototype to perform 3D BF, thanks to a compact design of the RIS control circuit. Indeed, it allows for the setting of 984 distinct voltages and associated phase-shifts, for the 984 distinct unit-cells of the RIS. This prototype, therefore individually configures with a distinct phase-shift, 70 times more unit-cells than in [5,6] and almost 20 times more than in [7] and [9].

For the first time, we also show how to produce a new radio type of equipment with minimum energy footprint and environmental impact by refurbishing an existing type of equipment. Indeed, we built the presented RIS prototype (Fig. 2-b) from a Reflect Array antenna [10,11] (illustrated in Fig. 2-a) with 3D transmit BF capability. During this refurbishing step, we kept the unit-cells and the control circuit and removed the transmit RF chain.

The paper is organized as follows: Section II presents the RIS prototype and the 3D reflective BF RIS tuning scheme, and Section III presents experimental results.

## II. RIS Prototype and Tuning

### A. RIS Prototype

The prototype, illustrated in Fig.2-b) is composed of 984 unit-cells. The unit-cell is depicted in Fig. 3-a). The unit-cell size is 14.0 by 14.0 mm. The frequency band of the unit-cell is 5.15-5.35 GHz. The unit-cell is a patch with central post and annular slot. Four varactor diodes connect the two parts of the patch. The detailed design of the unit cell is given in [10,11]. The phase-shift of each unit-cell is controlled individually through its four varactors by applying a voltage to the varactors. Fig. 2-c) illustrates the compact electronic control circuit allowing for an independent control of each of the 984 unit-cells. To achieve such challenge, the control has been split into four different circuits, each controlling a distinct quarter of 246 different unit-cells.

Note that, in addition to altering the phase-shift (which is the intended consequence), the voltage also slightly changes the unit-cell radiation pattern (which is a non-intended consequence), as illustrated in Fig. 3-b). Such behaviour is typical from varactor-based RIS prototypes [9].

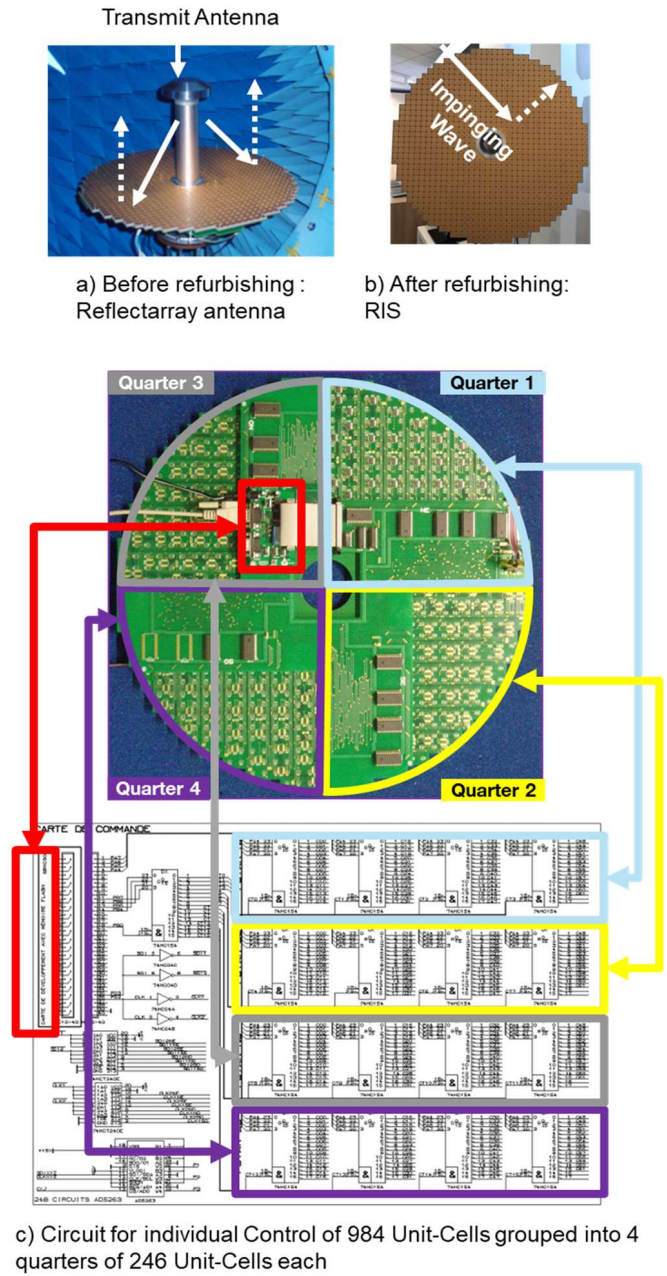

a) Before refurbishing: Reflectarray antenna

b) After refurbishing: RIS

c) Circuit for individual Control of 984 Unit-Cells grouped into 4 quarters of 246 Unit-Cells each

Fig. 2. Low footprint RIS prototype building

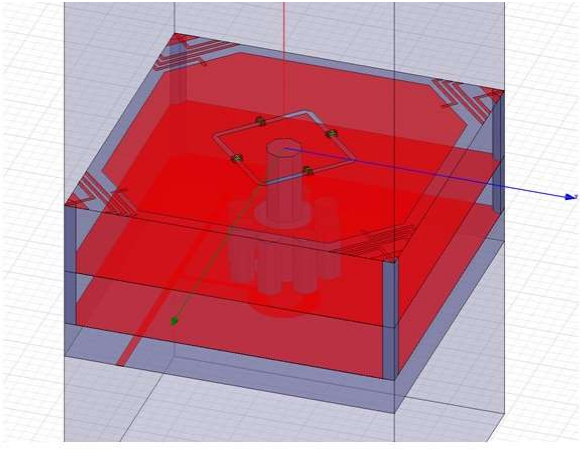

a) Unit-Cell

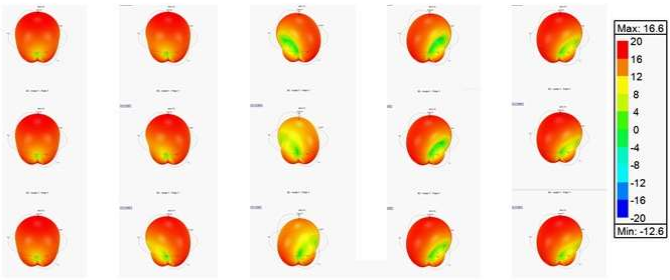

b) Diagram at 5.25 GHz for various voltages

Fig. 3. Unit-Cell (a) structure and (b) antenna diagrams.

Regarding the energy consumption of the RIS prototype, each unit-cell alone (to be switched to a given configuration) needs a femto Ampere of current and up to 5 Volts, i.e., up to 5 femto Watt per unit-cell. As the RIS has near to 1000 unit-cells, the total energy consumption (due to unit-cell varactors) is in the order of picowatt. To this energy consumption, one must add the energy necessary to control the circuit to send orders (i.e., in the micro-controller to generate the voltages in the digital domain and in the DACs to convert the voltages in the digital domain into the analog domain, etc.). With an integrated design of the control circuit, it is expected that the total RIS consumption remains in the order or even below 1 watt.

*B. RIS Prototype tuning for reflected BF*

The phase-shift of each unit-cell is computed individually as follows. The RIS is assumed to be in boresight of the source. Let $\theta$ and $\phi$ be the azimuth and elevation angles of the RX with respect to the RIS (as illustrated in Fig. 1-b). Let $f$ be the carrier frequency of the TX signal, $c$ be the light velocity and $\lambda = c/f$ be the wavelength. Let $\vec{u}$ be the vector between the center of the TX and the unit-cell $n$. Let $\vec{v}_n$ be the unitary vector in the RIS-RX direction (assuming RX is far from RIS). Let $\vec{w}_n$ be the vector between the RIS center and unit-cell $n$. The phase-shift due to the propagation delay due to the propagation between the TX and the RX after bouncing on the RIS is given by:

$$\alpha_n \sim 2\pi \frac{\|\vec{u}_n\| + \vec{w}_n \cdot \vec{u}}{\lambda}.$$

The phase-shift $\beta_n$ of the unit-cell is computed to compensate for the phase-shift of propagation $\alpha_n$, and ensure that all signals arrive in phase, coherently, at RX, as follows:

$$\beta_n = -\alpha_n.$$

Finally, the voltage to apply to the varactors of the unit-cell is derived, using the phase to voltage law illustrated in Fig. 4, which has been characterized experimentally [10,11].

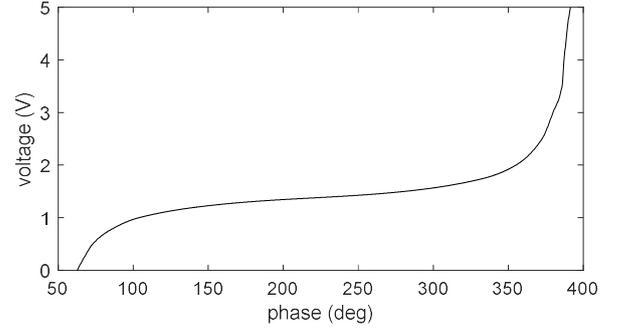

Fig. 4. Unit-cell voltage versus phase shift law

### III. EXPERIMENTS

Two different experiments are conducted indoors:

- Experimental set-up 1 assesses the ability of the RIS to alternate between two distinct directions,
- Experimental set-up 2 assesses the ability of the RIS to alternate between three distinct directions.

Sub-section A describes the TX and RX antennas, and Sub-sections B and C present the results of set-ups 1 and 2, respectively.

*A. TX and RX antennas*

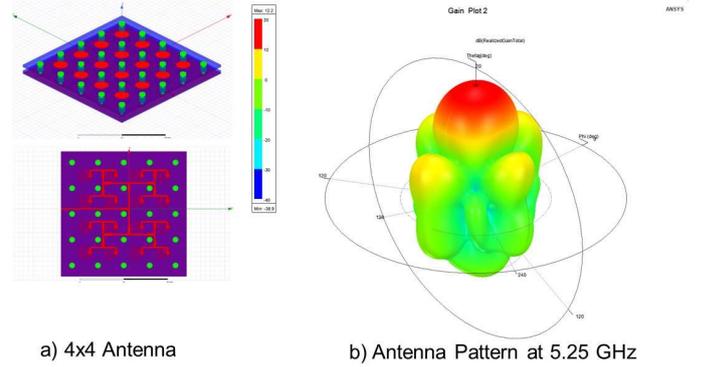

a) 4x4 Antenna            b) Antenna Pattern at 5.25 GHz

Fig. 5. TX Antenna

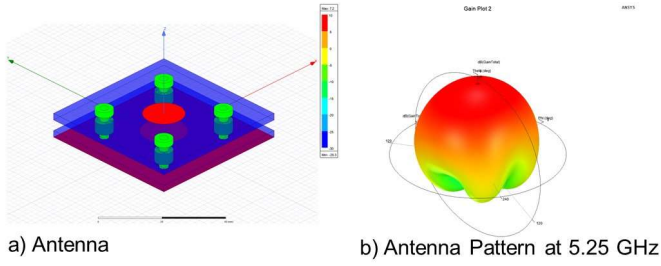

Fig. 6. RX Antenna

Experiments are conducted with the 4x4-port antenna illustrated in Fig. 5 for TX and the patch antenna illustrated in Fig. 6 for RX1, RX2, and RX3. The two types of antennas are very directional. The Tx antenna emits an orthogonal frequency division multiplex (OFDM) signal at f=5.25 GHz, with a phase shift keying (PSK) modulation with eight states, and a bandwidth of 60 MHz.

### B. Experimental Set-Up 1: 2 Distinct Reflected Beams

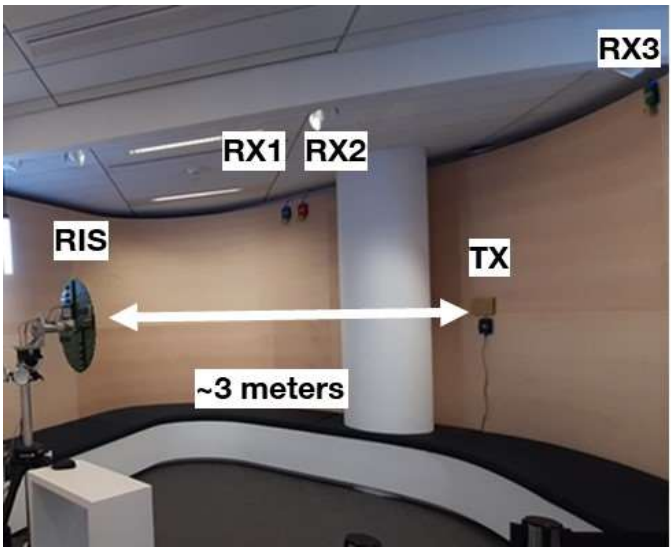

Fig. 7. Test Set-Up 1

Fig. 7 illustrates the experimental environment. The distance between TX and RIS is around 3.2 meters. RX1, RX2, and 3 are at similar distances from the RIS but at different elevations from the RIS than the TX. Also, as the TX and the RXs have directional antennas, the direct propagation between the TX and the RXs is very weak. By putting the RIS in the boresight of the TX, we expect to be able to reach the RXs, with some tuning of the phase-shifts. RX1 and RX2 are close to each other. We, therefore, expect to observe similar performance for these two RXs. Finally, RX3, on one side, and RX1 and RX2, on the other side, are separated by an obstacle. We, therefore, expect some contrast in the received powers by RX1 and RX2 on one side and RX3 on the other side.

The RIS is tested for two different configurations: Configuration A, aiming at reflecting towards RX1 and RX2, and Configuration B, targeting RX3. The RIS tunings for Configuration A and Configuration B are given in Table I.

TABLE I. RIS TUNINGS FOR SET-UP 1

| Configuration | $\theta(°)$ | $\varphi(°)$ |
|---|---|---|
| A (RX1,2) | $\theta_A = 27$ | $\varphi_A = 140$ |
| B (RX3) | $\theta_B = 25$ | $\varphi_B = 40$ |

Fig. 8-a) illustrates the voltages of the unit-cells for each configuration. Fig. 8-b) illustrates the expected reflected radiation pattern, of the RIS for each configuration, based on simulations. The expected contrast between the main lobe and side lobes is in the order of 20 to 35 dBs.

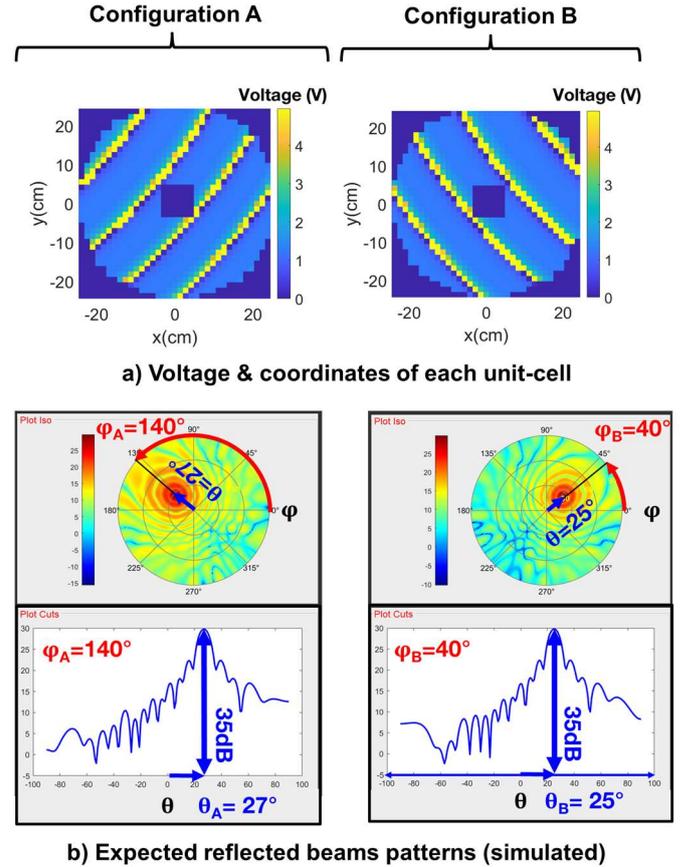

Fig. 8. RIS tunings and Simulated Reflected Beams for Set-Up 1

Fig. 9 and Fig. 10 illustrate the received signals at RX1, RX2, and RX3, in terms of power spectral densities and constellations, for configurations A and B, respectively. As expected, RX1 and 2 experience similar received power, whatever the configuration. We also observe that RX3, on one side, and RX1 and RX2, on the other side, experience different received powers. The observed contrast in power is around 10 dB for both configurations, as illustrated by Fig. 9 and Fig. 10. The observed contrast is lower than the simulated reflected beamforming gain of Fig. 8. This is in part due to the multipath propagation. This is confirmed by the constellations observed on the RX, which is not the target. Indeed, such constellations are spread over the complex real and imaginary domains due to channel time-frequency diversity. Finally, even though the TX and RX antennas are directional, the direct TX-RX propagation path is not negligible. Therefore, the received signals are not only due to the RIS reflected beam; they also depend on the TX-RX direct component. As the direct component is fixed, whatever the configuration of the RIS, it decreases the contrast in received power.

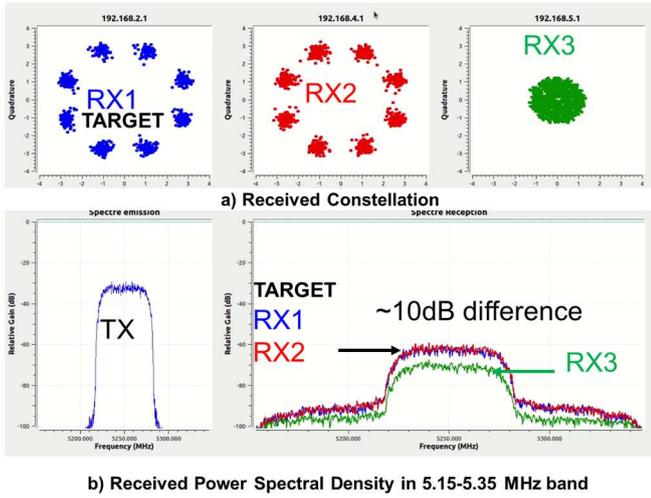

Fig. 9. Measured Received signals for configuration A

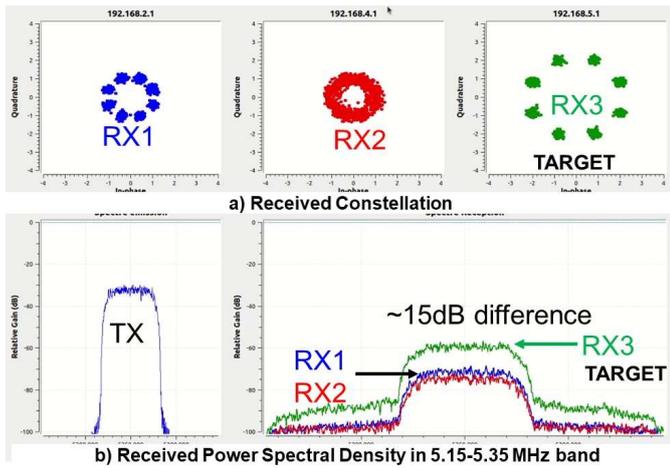

Fig. 10. Measured Received signals for configuration B

## C. Experimental Set-Up 2: 3 distinct reflected beams

The experimental environment is illustrated in Fig. 11. In this set-up, RX1, RX2, and RX3 are positioned in distinct locations. The RIS has three different configurations, A, B, and C targeting RX1, RX2, and RX3, respectively. The distances between the nodes and the RIS, and the angles, have similar values to the ones of set-up 1 (i.e., below 4 meters and within a cone of +/-45° around the boresight of the RIS-TX axis).

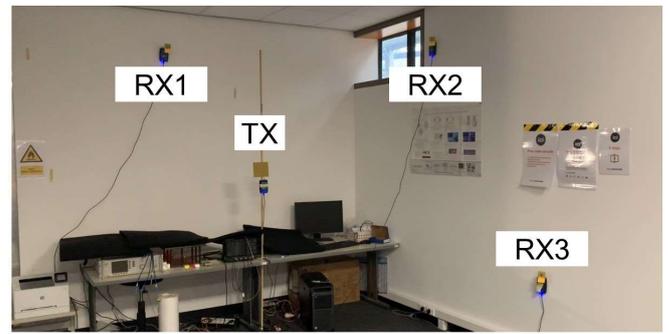

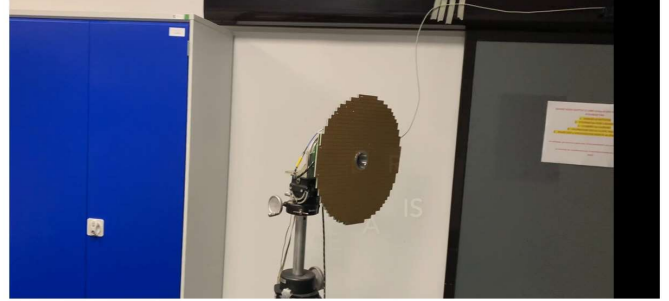

Fig. 11. Test set-up 2

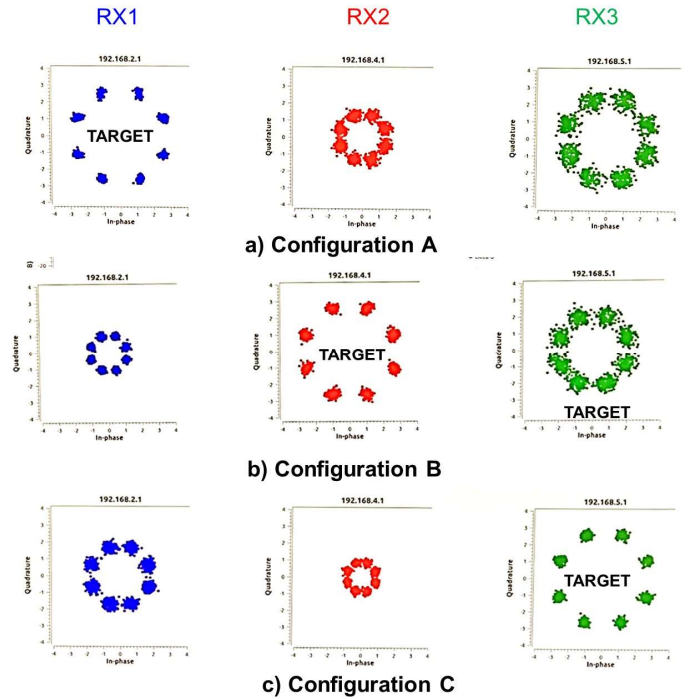

Fig. 12. Measured received constellations for set-up 2.

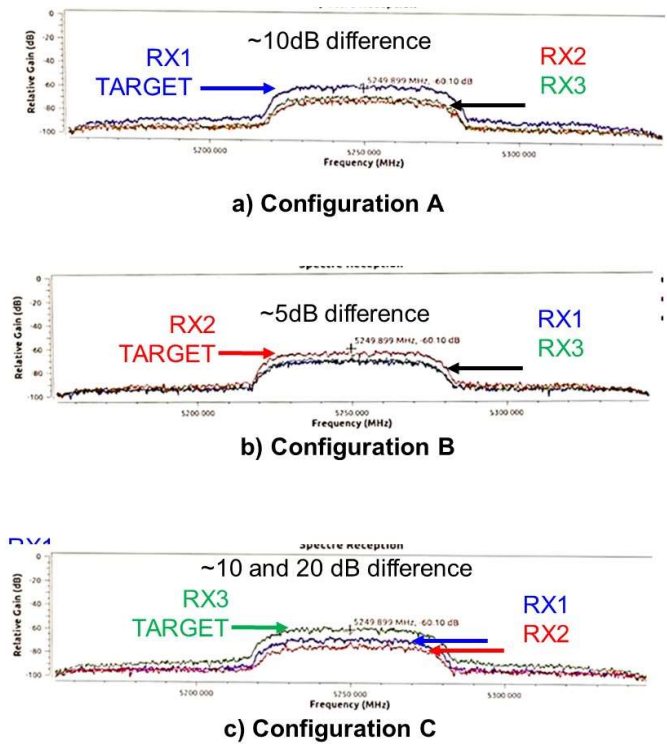

Fig. 13. Measured received power spectral density in the 5.15-5.70 MHz band, for set-up 2.

*D. Videos of experiments*

[12] and [13] are videos of experiments in set-up 2 and in similar conditions, respectively. They illustrate the role of each of the four groups of 246 unit-cells. Indeed, in these videos, the RIS switches from one configuration to the other, through four successive steps. During each step, the RIS updates the tuning of only one of the four groups of 246 unit-cells illustrated in Fig. 2, at a time. One can therefore observe the received power at the new target and the previous target RX, gradually increase and gradually decrease, in four successive steps.

## IV. CONCLUSION

In this paper, we present the first experimental results of 3D reflected beamforming with a varactor-based sub6GHz reconfigurable intelligent surface. This is achieved thanks to a compact control circuit addressing and configuring individually 984 distinct unit-cells with 984 distinct voltages, hence, 17 to 70 times more distinct voltages than in current state-of-the-art to our knowledge. We also show that a RIS can be built with minimum environmental footprint impact, by refurbishing a reflectarray antenna.

ACKNOWLEDGMENT

The authors are grateful for the support from Micro Ondes Antennes Recherche et Development (MARDEL) and the Centre de Recherche Mutualisé sur les Antennes (CREMANT). This work has been partially performed in the framework of the H2020 project RISE-6G under grant 101017011.